\documentclass[reprint,twocolumn,prl]{revtex4}
\usepackage[T1]{fontenc}
\usepackage[latin9]{inputenc}
\usepackage{color}
\usepackage{amsmath}
\usepackage{amssymb}
\usepackage{graphicx}
\usepackage{esint}
\usepackage{amssymb}

\makeatletter \@ifundefined{textcolor}{} {
 \definecolor{BLACK}{gray}{0}
 \definecolor{WHITE}{gray}{1}
 \definecolor{RED}{rgb}{1,0,0}
 \definecolor{GREEN}{rgb}{0,1,0}
 \definecolor{BLUE}{rgb}{0,0,1}
 \definecolor{CYAN}{cmyk}{1,0,0,0}
 \definecolor{MAGENTA}{cmyk}{0,1,0,0}
 \definecolor{YELLOW}{cmyk}{0,0,1,0}
 }

\begin{document}

\title{Comment on ``Trouble with the Lorentz Law of Force:
Incompatibility with Special Relativity and Momentum
Conservation''}
\author{F. De Zela}
\address{Departamento de Ciencias, Secci\'{o}n F\'{i}sica,
Pontificia Universidad Cat\'{o}lica del Perú, Ap. 1761, Lima,
Peru.}

\maketitle

Mansuripur claims that the Lorentz law of force must be abandoned
because it violates relativity \cite{Mansuripur}. To show this, he
considers a point
charge $q $ and a point magnetic dipole $m_{0}\widehat{\mathbf{x}}^{\prime }$%
, both being at rest in a reference system $x^{\prime }y^{\prime }z^{\prime }
$ that moves with constant velocity $\mathbf{V}=V\widehat{\mathbf{z}}$
relative to a second system $xyz$. The magnetization of the point dipole is
taken as $\mathbf{M}^{\prime }(\mathbf{r}^{\prime },t^{\prime })=m_{0}%
\widehat{\mathbf{x}}^{\prime }\delta (x^{\prime })\delta (y^{\prime })\delta
(z^{\prime }-d)$ in the $x^{\prime }y^{\prime }z^{\prime }$-system. By
Lorentz transforming $\mathbf{M}^{\prime }$ one obtains $\mathbf{M}$ and $%
\mathbf{P}$ as the components of the second-rank tensor describing the point
dipole in the $xyz$-system. Using the Lorentz force expression $\mathbf{F}(%
\mathbf{r},t)=\rho _{bound}\mathbf{E}+\mathbf{J}_{bound}\times \mathbf{B}$,
it can be shown that the net force $\int \mathbf{F}(\mathbf{r},t)d^{3}r$ on
the point dipole is zero, while the net torque $\mathbf{T}=\int \mathbf{r}%
\times \mathbf{F}(\mathbf{r},t)d^{3}r=(Vqm_{0}/4\pi d^{2})\widehat{\mathbf{x}%
}$. Mansuripur argues that the appearance of a nonzero torque in the $xyz$%
-system -- in the absence of a corresponding torque in the
$x^{\prime }y^{\prime }z^{\prime }$-system -- is sufficient proof
of the inadequacy of the Lorentz law. We believe that Mansuripur's
conclusion is premature, for the following reasons.

First of all, the mere appearance of a torque in only one of two inertial
systems does not violate relativity. For example, Jackson \cite{Jackson1}
has recently discussed the case of a point charge $q$ moving in the central
field of a second charge $Q$ that remains fixed in an inertial system, the $%
x^{\prime }y^{\prime }z^{\prime }$-system, say. As seen from the $x^{\prime
}y^{\prime }z^{\prime }$-system, charge $q$ experiences no torque and moves
along the straight line that joins it with $Q$. However, when seen from the $%
xyz$-system -- which uniformly moves with respect to the $x^{\prime
}y^{\prime }z^{\prime }$-system -- a torque appears, causing a continuous
change in the angular momentum of $q$. Jackson shows that there is no
paradox here. Everything is in perfect accordance with relativity. It may
exist a torque causing a change of angular momentum in one inertial frame,
while there is no angular momentum and no torque in the other frame. Thus,
the appearance of a nonzero torque in only one of two inertial systems is
not by itself a contradiction. Coming back to Mansuripur's case, if we had
obtained that the net force is zero in one system and nonzero in the other,
then we would have been faced with a paradox. But the paradox would have not
consisted in having two different values for the force. The paradox would
have arisen from the observable consequences of this fact. Indeed, in such a
case the particle's motion would have appeared as accelerated motion in one
inertial frame and as uniform motion in the other. And this would be in
conflict with relativity. Can we similarly reason with the torque? The
answer is not. Indeed, as we have a vanishing net force, the torque is an
internal one with no effect on the body's center of mass motion. The
torque's sole effect must be a rotation of the body around some axis passing
through it. However, if the body is a point, such a rotation becomes
meaningless. As already said, the fact that a torque appears in the $xyz$%
-system and not in the $x^{\prime }y^{\prime }z^{\prime }$-system is not by
itself a paradox \cite{Jackson1}. Moreover, in Mansuripur's case such a fact
does not even have any observable consequences, as in Jackson's case. Thus,
it does not imply a contradiction, like motion that is accelerated in one
inertial frame and uniform in the other. Here we are faced with a torque
having no observable effects that might violate relativity. Instead, such a
result rather hints at some possible inconsistencies that may arise when
using models of ``point'' magnetic dipoles and the like. We analyze this
issue in more detail in what follows.

To begin with, quantities like $\mathbf{P}$ and $\mathbf{M}$ serve to
describe electromagnetic properties of bulk material. A magnetic dipole
represents the lowest order multipole contribution from a divergenceless
current distribution $\mathbf{J}$ within a sample \cite{Jackson}: $\mathbf{M}%
\propto \int \mathbf{r\times J(r)}d^{3}r$. In classical electrodynamics,
\emph{point} dipoles are no more than convenient idealizations. Dipoles do
not appear as fundamental quantities, like charges and fields; but as
derived quantities, owing their existence to charge and current
distributions. We may certainly consider very small pieces of material whose
electromagnetic properties are described by smooth functions of space and
time, e.g., $\mathbf{P}$ and $\mathbf{M}$. But there is no sensible physical
picture associated with a true \emph{pointlike} object carrying \emph{%
internal} currents. Idealized entities correspond to real entities
whose finite extent we may neglect after having fixed the accuracy
of our calculations. We may then employ mathematical tools like
Dirac's $\delta $ to describe our idealization of the true
physical objects. As well known, Dirac's $\delta $ is a
distribution, not a function. As such, it may be used to describe
idealized entities having no extent. We should bear in mind that
with the help of Dirac's delta we may derive properties that
belong to idealized entities, but not necessarily to real ones
\cite{fn}. If we give ourselves carte blanche in the usage of
Dirac's delta, we risk loosing the self-consistency of our
physical model, thereby obtaining contradictory results. As an
illustration of this, let us consider instead of the Lorentz force
law, the simpler expression for the electric force-density: $\rho
\mathbf{E}$. Such an expression follows from the very definition of $\mathbf{%
E}$ as force per unit charge, and of $\rho $ as charge per unit volume. We
have thus an expression which is valid by definition. Now, following a
similar reasoning and employing the same techniques as in \cite{Mansuripur},
we can get contradictory results. Let us see how. Consider the
point-dipole's polarization
\begin{equation*}
\mathbf{P}(\mathbf{r},t)=\gamma V\varepsilon _{0}m_{0}\widehat{\mathbf{y}}%
\delta (x)\delta (y)\delta \left[ \gamma (z-Vt)-d\right] .
\end{equation*}%
It relates to the charge density through
\begin{equation*}
\rho _{bound}=-\mathbf{\nabla }\cdot \mathbf{P}=\gamma V\varepsilon
_{0}m_{0}\delta (x)\delta ^{\prime }(y)\delta \left[ \gamma (z-Vt)-d\right] .
\end{equation*}%
Hence, the total charge of the dipole is
\begin{eqnarray*}
\int \gamma V\varepsilon _{0}m_{0}\delta (x)\delta ^{\prime
}(y)\delta \left[ \gamma (z-Vt)-d\right] d^{3}r=\\
=\int V\epsilon _{0}m_{0}\delta ^{\prime }(y)dy=-d(V\epsilon
_{0}m_{0})/dy=0,
\end{eqnarray*}%
as it should. Let us assume for a moment that we are using Dirac's deltas
for describing not the idealized but the real entities. We would be then
describing a neutral and \emph{pointlike} particle. Now, structureless,
i.e., truly pointlike particles carrying no electric charge do not couple to
the electromagnetic field. If we apply an electric field $\mathbf{E}$ to
such a neutral \emph{pointlike} particle, then the field should exert no
force upon it. However, if we calculate this force for the Lorentz
transformed Coulomb field
\begin{equation*}
\mathbf{E}=\left( \frac{\gamma q}{4\pi \varepsilon _{0}}\right) \frac{x%
\widehat{\mathbf{x}}+y\widehat{\mathbf{y}}+(z-Vt)\widehat{\mathbf{z}}}{%
\left( x^{2}+y^{2}+(z-Vt)^{2}\right) ^{3/2}},
\end{equation*}%
we obtain $\int \left( -\mathbf{\nabla }\cdot \mathbf{P}\right) \mathbf{E}(%
\mathbf{r},t)d^{3}r=\left( \gamma Vm_{0}q/4\pi d^{3}\right) \widehat{\mathbf{%
y}}$. The same contradiction arises by considering a situation similar to
that in \cite{Mansuripur} but starting with a dipole $\mathbf{P}^{\prime
}=p_{0}\widehat{\mathbf{y}}^{\prime }\delta (x^{\prime })\delta (y^{\prime
})\delta (z^{\prime }-d)$ in place of $\mathbf{M}^{\prime }$. Such a
contradiction does not invalidate the expression for the force-density, $%
\rho \mathbf{E}$, which follows -- as already said -- from the
very definitions of $\mathbf{E}$ and $\rho $. What the
contradiction signals is that we cannot use $\mathbf{P}$ in
connection with a true point-particle. Endowing a true
point-particle with polarization and/or magnetization properties
makes no sense in the framework of classical electrodynamics. We
can certainly deal with a very small piece of matter that is
polarized and/or magnetized, neglect its extension and describe it
as a point. Dirac's delta can then be used as an appropriate tool.
But it is the right tool only as long as we deal with the
idealized model. Not every result that can be obtained with the
help of such a tool necessarily applies for the real entities. If
we do not distinguish the idealized model from its real
counterpart, we can arrive at contradictory results, as
illustrated by the above cases. As shown below, other rebuttals of
Mansuripur's ``paradox''
\cite{Vanzella,Griffiths,Saldanha,Brachet} illustrate this point
as well.

Like in Jackson's example \cite{Jackson1}, there are other
apparent paradoxes that on closer look rather confirm the validity
of classical electromagnetism, whenever one deals with cases lying
within the scope of this theory. For example, if we properly
handle a small current loop as the source of the magnetic moment
$m_{0}\widehat{\mathbf{x}}^{\prime }$, everything fits. This has
been shown by Saldanha \cite{Saldanha}, who essentially reproduces
a ``paradox'' discussed in Griffiths's well-known electrodynamics
textbook \cite{Griffiths2}, and by Griffiths and Hnizdo
\cite{Griffiths} in their rebuttal to \cite{Mansuripur}. However,
as soon as we go from the very small to the \emph{pointlike}, the
shortcomings of classical electromagnetism begin to appear. To see
this, let us consider in more detail the solution of Mansuripur's
``paradox'' proposed by Griffiths and Hnizdo \cite{Griffiths}.
These authors consider two different models for Mansuripur's
magnetic dipole. One model is a ``Gilbert dipole'' (two magnetic
monopoles) and the other is an ``Ampere dipole'' (an electric
current loop). Now, the first model lies beyond classical
electromagnetism, as it makes use of magnetic monopoles and the
corresponding modification of the Maxwell and the Lorentz
equations. If we want to prove that a given example does not lead
to a true paradox within some theoretical framework, we cannot go
beyond that framework. A paradox tries to show the inconsistency
of some theoretical construction. By ``solving'' it with tools
taken from outside this theoretical construction, we are doing
nothing but confirming the shortcomings of that framework, which
is precisely what the paradox intended to do. We should thus
consider only the ``Ampere dipole''. As shown in \cite{Griffiths},
by considering from the perspectives of two inertial systems a
current loop and a charge in relative rest, no paradox arises. The
appearance of a torque and the corresponding changing angular
momentum in one system and not in the other is in total agreement
with relativity. The concept of ``hidden momentum'' explains all
the observed phenomena \cite{Griffiths,Shockley,Furry,Coleman}.
Though this concept is somewhat controversial, there are cases
like those addressed by Griffiths and Hnizdo \cite{Griffiths}, in
which hidden momentum is an incontrovertible relativistic effect
that must be taken into account. In all these cases one deals with
finite samples of matter. Inconsistencies appear when one goes to
the pointlike limit. In the case of an ``Ampere
dipole'', for instance, the hidden momentum is given by $%
\mathbf{p}_{h}=\mathbf{m}\times \mathbf{E}/c^{2}$ in the units and
notation of \cite{Griffiths}. Considering a case analogous to
Mansuripur's, $\mathbf{p}_{h}$ points perpendicularly to the
relative velocity and hence remains unchanged when going from the
$x^{\prime }y^{\prime }z^{\prime }$-system to the $xyz$-system.
The associated hidden angular momentum
$\mathbf{L}_{h}=\mathbf{r}\times \mathbf{p}_{h}$ has, in
turn, a rate of change given by \cite{Griffiths} $d\mathbf{L}_{h}/dt=\mathbf{%
v}\times \left( \mathbf{m}\times \mathbf{E}\right) /c^{2}$, as seen from the $%
xyz$-system. This is just the value obtained for the torque
exerted on the current loop. So, everything appears to be
consistent in this case. Inconsistencies show up when we deal with
pointlike objects, as Mansuripur did. Grifiths and Hnizdo address
this case as well, but in a way that lacks the self-consistency of
their previous treatment.
Indeed, when dealing with Mansuripur's case, they keep using the expression $%
\mathbf{p}_{h}=\mathbf{m}\times \mathbf{E} /c^{2}$ for the hidden
momentum. This is a highly questionable procedure. Such an
expression for $\mathbf{p}_{h}$ is obtained under a series of
assumptions which do not hold for the case at hand. Let us briefly
review how the hidden momentum appears
\cite{Hnizdo,Furry,Coleman}. One considers the
energy-momentum tensor $T^{\mu \nu }$ of a system of particles (a ``body''). It satisfies the equation $%
\partial T^{\mu \nu }/\partial x^{\nu }=f^{\mu }$, with $f^{\mu }=\left(
\mathbf{f}\cdot \mathbf{v}/c,\mathbf{f}\right) $ the four-vector
representing the force-density that acts within the body. The $\mu
=0$ component of the above equation reads $\partial
u/\partial t+\mathbf{\nabla }\cdot (c^{2}\mathbf{g})=\mathbf{f}\cdot \mathbf{%
v}$, with $u=T^{00}$ being the energy density and $g_{i}=T^{0i}/c$
the three components of the momentum density $\mathbf{g}$. Using
these results one can prove that there is a ``hidden momentum''
$\mathbf{p}_{h}:=\int \mathbf{g}d^{3}r$, which shows up even when
the body's elements move in a stationary way ($\partial u/\partial
t=0$) and the body's center of mass remains at rest. In the
stationary case, $\mathbf{p}_{h}=-c^{-2}\int
\mathbf{r}(\mathbf{f}\cdot\mathbf{v})d^{3}r$. When the force
density is given by $\mathbf{f}=-\rho \mathbf{\nabla }\phi $, one
obtains \cite{Hnizdo} $\mathbf{p}_{h}=-c^{-2}\int \phi \rho
\mathbf{v}d^{3}r$. And for the special case of
a uniform, \emph{external} electric field $\mathbf{E}$ one derives \cite{Hnizdo} (using $\phi =-%
\mathbf{E}\cdot \mathbf{r}$) the alternative expression $%
\mathbf{p}_{h}=\mathbf{m}\times \mathbf{E}/c^{2}$, with $%
\mathbf{m}=(1/2)\int \mathbf{r\times j}d^{3}r$ denoting the magnetic moment
associated to the current density $\mathbf{j}=\rho \mathbf{v}$.

Although the above results can be applied to a small current loop,
it is highly questionable that we may keep applying them when
dealing with a pointlike object. For, first, the force density
$f^{\mu }$ referred to above is the total one acting on a given
element of the body \cite{Hnizdo}. That is, it includes external
forces as well as the forces originating from the rest of the
body. We can neglect internal contributions only under special
assumptions. Generally, any such assumptions become invalid when
we let the body shrink to sufficiently small dimensions. Classical
electromagnetism predicts that below some distance internal forces
turn out to be larger than any
external forces. But in addition to this objection, there is a second one. We have seen that $%
\mathbf{p}_{h}=\mathbf{m}\times \mathbf{E} /c^{2}$ applies for an
external force-density given by $\mathbf{f}=-\rho \mathbf{\nabla
}\phi $. In Mansuripur's case, however, $\mathbf{f}$ is given (in
the notation of \cite{Griffiths}) by $\mathbf{f}=-qm_{0}vd\left(
4\pi \epsilon _{0}c^{2}R^{3}\right) ^{-1}\delta (x)\delta ^{\prime
}(y)\delta \left[ z-Vt-d/\gamma \right] \widehat{\mathbf{z}}$,
with $R=\left( x^{2}+y^{2}+(z-Vt)^{2}\right) ^{1/2}$. Using this
$\mathbf{f}$ in
$\mathbf{p}_{h}=-c^{-2}\int%
\mathbf{r}(\mathbf{f}\cdot\mathbf{v})d^{3}r$ we do not obtain the
desired results. Thus, as stressed by Boyer \cite{Boyer}, hidden
momentum can lend itself to ``explanations of dubious validity or
outright error which avoid needed discussions of energy and
momentum flow''. Such discussions can be carried out when
dealing with models that fall in line with the tenets of classical
electromagnetism. This is not to say that classical
electromagnetism is a complete, self-consistent theory of
electromagnetic phenomena. Even situations that are much simpler
than the ones we have discussed so far, do require that we resort
to quantum mechanics.
Consider for example a point charge $q$ sitting at rest at the origin $%
O$ of the $xyz$-system. In accordance with Maxwell's equations,
$q$ is the source of a purely electrostatic field, the Coulomb
field. Consider now the same situation as seen from the
perspective of the $x^{\prime }y^{\prime }z^{\prime }$-system,
whose origin coincides with $O$ while its axes rotate with respect
to the $xyz$-system. Although the exact \emph{space-time}
transformation linking the two \emph{reference} frames (in
contrast to the \emph{coordinate systems}, $t^{\prime }x^{\prime
}y^{\prime }z^{\prime }$ and $txyz$) is unknown, we can certainly
take for granted that classical electromagnetism predicts that in
the rotating frame a magnetic field will be observed, together
with an electric field. In other words, in the $x^{\prime
}y^{\prime }z^{\prime }$-system $q$ would be the source of an
electric and a magnetic field. Now, given a true \emph{point}
charge sitting at rest, how could we tell whether it is rotating
or not? We cannot have a consistent model of elementary charges at
rest, unless we provide all of them with the property of being
sources of both electric and magnetic fields. This is precisely
what quantum mechanics does by endowing the electron -- and other
elementary particles -- with a spin. The quantum mechanical
electron is thus a source of both electric and magnetic fields. At
the same time, even being at rest the pointlike electron can be
set in ``rotation'' by an external magnetic field. But now
``rotation'' acquires a physical meaning: the observable spin
dynamics.

In conclusion, Mansuripur's claim that the Lorentz law should be abandoned
seems to be unsubstantiated. The mere existence of a torque that appears in
one inertial frame and not in the other does not violate relativity, as it
has been illustrated by Jackson's example \cite{Jackson1}. Mansuripur's
results would rather illustrate the limits of classical models, when it
comes to describe some electromagnetic features.


\begin{thebibliography}{99}
\bibitem{fn} In physics, as in mathematics,
a limiting procedure can bring us out from a given framework. E.g.,
any finite sum of the form $r_{n}=4-4/3+4/5-\ldots
(-1)^{n}4/(2n+1) $ is a rational number ($r_{n}\in\mathbb{Q}$).
However, $\lim_{n\rightarrow\infty} r_{n}=\pi \notin \mathbb{Q}$.
Similarly, Dirac's delta, being not a regular function but a
distribution, is a limiting case of regular Gaussian functions.
Whether such qualitative changes are relevant or not for the
corresponding physical models, is something that must be decided in each
particular case.
\end{thebibliography}
\end{document}